\documentclass[prl,twocolumn,nofootinbib]{revtex4-1}
\usepackage{graphicx}
\usepackage{amssymb,amsmath,enumerate}
\usepackage{color}
\usepackage{bm}

\newcommand{\be}{\begin{eqnarray}}
\newcommand{\ee}{\end{eqnarray}}

\begin{document}

\title{Unifying size-topology relations in random packings of dry adhesive polydisperse spheres}

\author{Wenwei Liu$^1$, Sheng Chen$^2$, Chuan-yu Wu$^1$, Shuiqing Li$^2$}

\affiliation{$^1$Department of Chemical and Process Engineering, University of Surrey, Guildford, GU2 7XH, UK \\
$^2$Key Laboratory for Thermal Science and Power Engineering of Ministry of Education, Department of Energy and Power Engineering, Tsinghua University, Beijing 100084, China
}

\begin{abstract}

We study the size-topology relations in random packings of dry adhesive polydisperse microspheres with Gaussian and lognormal size distributions through a geometric tessellation. We find that the dependence of the neighbour number on the centric particle size is always quasilinear, independent of the size distribution, the size span or interparticle adhesion. The average local packing fraction as a function of normalized particle size for different size variances is well regressed on the same profile, which grows to larger values as the relative strength of adhesion decreases. As for the local coordination number-particle size profiles, they converge onto a single curve for all the adhesive particles, but will gradually transfer to another branch for non-adhesive particles. Such adhesion induced size-topology relations are interpreted theoretically by a modified geometrical ``granocentric'' model, where the model parameters are dependent on a dimensionless adhesion number. Our findings, together with the modified theory, provide a more unified perspective on the substantial geometry of amorphous polydisperse systems, especially those with fairly loose structures.

\end{abstract}

\date{\today}

\maketitle

\section{Introduction}

As the most fundamental property, size polydispersity is omnipresent in almost all realistic particle systems and industrial applications, including glasses, emulsions, colloids, granular materials, as well as geotechnical and process engineering \cite{Bernal59, Parisi10, Torquato10, Jorjadze11, Coniglio04, Andreotti13, Mitchell05, Xiao18}. It is well known to have a great impact on the material's macroscopic behaviour, such as the packing fraction and strength properties of materials \cite{Voivret09, Shaebani12, Desmond14, Estrada16}. In highly polydisperse granular materials, smaller particles can pack more efficiently by either fitting into the voids between large particles or by layering against them \cite{Desmond14}. Therefore, understanding how particle packing is affected by size polydispersity has been a substantial area of study in recent years. In parallel with the large amount of literature focusing on monodisperse sphere packings, many experimental, theoretical, and numerical works have been devoted to the analysis of polydisperse systems \cite{Voivret09, Shaebani12, Desmond14, Zhang15, Estrada16}. Particular interest has been drawn to uncovering the underlying packing geometry and topology \cite{Yang02, Glotzer11, Newhall12}, which contributes to the in-depth understanding of jamming transition \cite{Ono02, OHern03, Majmudar07, Schroter07} as well as the optimization in designing particle-based materials \cite{Mitchell05}. Several theoretical approaches have been proposed to understand the local and global packing properties, ranging from geometric tilling \cite{Clusel09, Corwin10, Newhall11} to statistical mechanics of jammed matter \cite{Song08, Danisch10, Ciamarra12, Baule18, Biazzo09}. These theories quantify the randomness and fluctuations of the macroscopic properties through the probability distributions of local parameters, such as the coordination number, neighbour number and the cell volume. A governing principle that predicts the packing geometry is thus well established specifically for compressed and jammed configurations.

However, most packings of dry small micron-sized particles in nature are subject to adhesive forces as well \cite{Marshall14, Dominik97, Blum04, Kinch07}, where fairly loose but mechanically stable structures are observed. Despite that there have been a few investigations on such adhesive packings of monodisperse microspheres \cite{Yang00, Valverde04, Blum06, Martin08, Parteli14, Liu15, Liu16}, the influence of size polydispersity on adhesive packings has not been systematically addressed. The difficulty arises from the great challenges in the experimental measurements on the local packing properties, particularly for dry microparticle system. Moreover, as far as we know, there is barely no available theoretical interpretation for adhesive polydisperse packings. Therefore, numerical simulation with discrete element method (DEM) or molecular dynamics (MD) has become a powerful tool to take a deep insight into the adhesive packings \cite{Yang00, Brilliantov07, Martin08}. By means of the numerical approach, the local parameters in the packing can be easily attained and applied in the further analysis.

In this work, we study the adhesion induced size-topology relations in the adhesive loose packing of dry polydisperse microspheres through a geometrical tessellation. The packing structures are generated with a novel DEM framework that is specially developed for adhesive microspheres \cite{Marshall14, Li11}. The global and local packing fraction/coordination number are carefully measured and the neighbours of each particle in the packing are identified by the navigation map. The size-topology relations, which refers to the geometry in a local cell, are explained with a modified granocentric model \cite{Clusel09, Corwin10, Newhall11}.

\section{Methodology}

We generate a series of mechanically stable random packings, each of which consists of 5,000 polydisperse spheres, in a cuboid domain by means of discrete element simulation. Periodic boundary conditions are applied along the two horizontal directions to eliminate the wall effects, while gravity is set along the vertical direction. Particles are treated as deformable elastic spheres. The interparticle van der Waals adhesion is introduced by the surface energy $\gamma$, which is a fixed value $\gamma=15mJ/m^2$ in this study based on the atom force microscope measurements \cite{Heim99}. The Johnson-Kendall-Roberts (JKR) model is employed to account for normal contact force between the relatively compliant microspheres \cite{Johnson71}. Apart from the normal force, the tangential sliding forces, the rolling resistance and the twisting resistance are fully considered and included in the computational model, which are all approximated by the linear spring--dashpot--slider model \cite{Li11, Marshall14}.

Depending on the model used, the forces and torques exerting on each particle are given as,
\be
\label{Eq1}
\begin{split}
F_n &= 4F_C[(a/a_0)^3-(a/a_0)^{3/2}] + \eta_N {\bf v}_R \cdot {\bf n}, \\
F_s &= k_T[\int_{t_0}^t {\bf v}_R(\tau) \cdot {\bf t}_s d\tau] + \eta_T{\bf v}_R \cdot {\bf t}_s, \\
M_t &= \frac{k_Ta^2}{2}[\int_{t_0}^t {\bf \Omega}_t(\tau)d\tau] + \frac{\eta_Ta^2}{2}{\bf \Omega}_t, \\
M_r &= 4F_C(\frac{a}{a_0})^{3/2}[\int_{t_0}^t {\bf v}_L(\tau)d\tau],
\end{split}
\ee
where $F_n$, $F_s$, $M_t$, $M_r$ denote the normal force, tangential sliding force, twisting resistance torque and rolling resistance torque, respectively. $F_C=3\pi \gamma R$ is the critical pull-off force derived from the JKR theory \cite{Johnson71}, where R is defined as the effective radius between two contacting particles, $1/R=1/r_{p,i}+1/r_{p,j}$, and $a$ is the radius of contact area with $a_0$ at the equilibrium state. ${\bf v}_R \cdot {\bf t}_s$, ${\bf \Omega}_t$ and ${\bf v}_L$ are the relative sliding, twisting and rolling velocity at the contact point, respectively. ${\bf n}$ and ${\bf t}_s$ are the unit vectors in the normal and tangential direction, respectively. $\eta_N$ is the normal dissipation coefficient, and $k_T$ and $\eta_T$ are the tangential stiffness and dissipation coefficient, respectively.

According to Eq.~\ref{Eq1}, $F_s$, $M_t$, and $M_r$ will accumulatively increase as the relative sliding, twisting and rolling displacements increase. The tangential sliding force is governed by the Coulomb's friction law, where the sliding force becomes constant after reaching a limiting value $F_{s,crit}$ and particle will start to slide over each other. This limit in the presence of adhesion is given as \cite{Li11, Marshall14},
\be
\label{Eq2}
\begin{split}
F_{s,crit} &= \mu_f F_C |4(a/a_0)^3-(a/a_0)^{3/2} + 2|,
\end{split}
\ee
where $\mu_f$ is the sliding friction coefficient. Similarly, particles will irreversibly spin or roll over its neighbour when $M_t$ and $M_r$ reach their critical values $M_{t,crit}$ and $M_{r,crit}$, which are expressed in the presence of adhesion as,
\be
\label{Eq3}
\begin{split}
M_{t,crit} &= 3\pi aF_{s,crit}/16, \\
M_{r,crit} &= -4F_C(a/a_0)^{3/2}\theta_{crit}R. 
\end{split}
\ee
Here, $\theta_{crit}$ is the critical rolling angle, which is around $0.6\%-1.0\%$ according to the measurement by atomic force spectroscopy \cite{Sumer08}. More details about the DEM and the model parameters can be found in \cite{Li11, Yang13}.

The procedure for obtaining mechanically stable random packings is implemented by continuously dropping the particles into the domain with an initial velocity $U_0$, which resembles the random ballistic deposition approach \cite{Blum06}. A mechanically stable packing structure is obtained when all the particles are at rest after a sufficiently long time to dissipate all the kinetic energy. The particles are polydisperse with the mean radius $r_{p0}$ ranging from $2\mu m$ to $100\mu m$ and the normalized size variance $\sigma=\sigma_r/r_{p0}$ varying from $0$ to $0.4$. The height of the domain is $80r_{p0}$, while the length and width are $20r_{p0} \times 20r_{p0}$. It should be noted that in our present study, no flow field or electrostatic interactions are taken into account. Therefore, the packings are only subject to interparticle collision and gravity. It was previously reported that the relative strength of interparticle adhesion can be characterized by a dimensionless adhesion number, $Ad=\gamma /(\rho_p U^2 r_{p0})$ \cite{Li07}, which is defined as the ratio between the surface energy and particle inertia. The larger the $Ad$ is, the more adhesive the particle is. The adhesion number has been proved to be a well-defined dimensionless parameter in describing the dynamic behavior of adhesive particles, including random packings \cite{Liu15,Chen16} and fiber filtration \cite{Yang13}. Therefore, we use this adhesion number in our present study to address the influence of adhesion. Furthermore, we apply two different size distributions but with the same $r_{p0}$ and $\sigma$, i.e. the Gaussian distribution and lognormal distribution, in order to study the effects of the size distribution. All the parameters used in the simulation are summarized in Table~\ref{Table1}. Each simulation is run three times to guarantee a reliable average result.

\begin{table}[h]
\small
  \caption{\ Parameters used in DEM simulations.}
  \label{Table1}
  \begin{tabular*}{0.5\textwidth}{@{\extracolsep{\fill}}ccc}
    \hline
    Parameter & Value & Unit \\
    \hline    
    Mean particle radius ($r_{p0}$) & 2-100 & $\mu m$ \\
    Normalized variance ($\sigma$) & 0-0.4 & \\
    Particle number ($N$) & 5000 &  \\
    Particle mass density ($\rho_p$) & 2500 & $kg/m^3$ \\
    Surface energy ($\gamma$) & 15 & $mJ/m^2$ \\
    Domain length & $20r_{p0}$ & $\mu m$ \\
    Domain height & $80r_{p0}$ & $\mu m$ \\
    Gravity acceleration ($g$) & 9.81 & $m/s^2$ \\
    Initial velocity ($U_0$) & 0.5-2 & $m/s$ \\
    Friction coefficient ($\mu_f$) & 0.3 &  \\
    \hline
  \end{tabular*}
\end{table}

\begin{figure}
\begin{center}
\includegraphics[width=7.5cm]{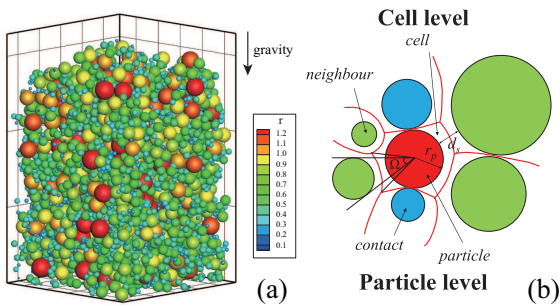}
\caption{\label{Fig1}(Colors online) Schematics of (a) the packing structure with $r_{p0}=10\mu m$ and $\sigma=0.4$, and (b) the navigation map.}
\end{center}
\end{figure}

Fig.\ref{Fig1}a shows a typical mechanically stable packing structure, where the color bar represents the normalized particle radius $r=r_p/r_{p0}$. The packing structure is then tessellated through the navigation map \cite{Richard01}, which is an extension of the Voronoi tessellation to polydisperse system, to obtain the local microstructure of each particle. In the navigation map, the space is partitioned into non-overlapping cells that are separated by hyperbolic surfaces. Each cell contains only one particle and all the points that are closest to its surface. Two particles are defined as neighbours if their corresponding cells share a common interface. Fig.\ref{Fig1}b displays the schematic of the navigation map in 2D, where the hyperbolic surface is reduced to hyperbola for better illustration. With the Monte Carlo method, we are able to estimate the cell volume $V_c$  and thus the local packing fraction is determined by $\phi_{local}=4\pi r_p^3/(3V_c)$. Then the local coordination number is decided by judging whether the distance between the two particles are smaller than the sum of their radii. It should be noted that the neighbours of a particle are not necessarily in touch with it, so that the coordination number is always less than or equal to the number of neighbours.

\section{Results and Discussions}

\begin{figure*}
\begin{center}
\includegraphics[width=18cm]{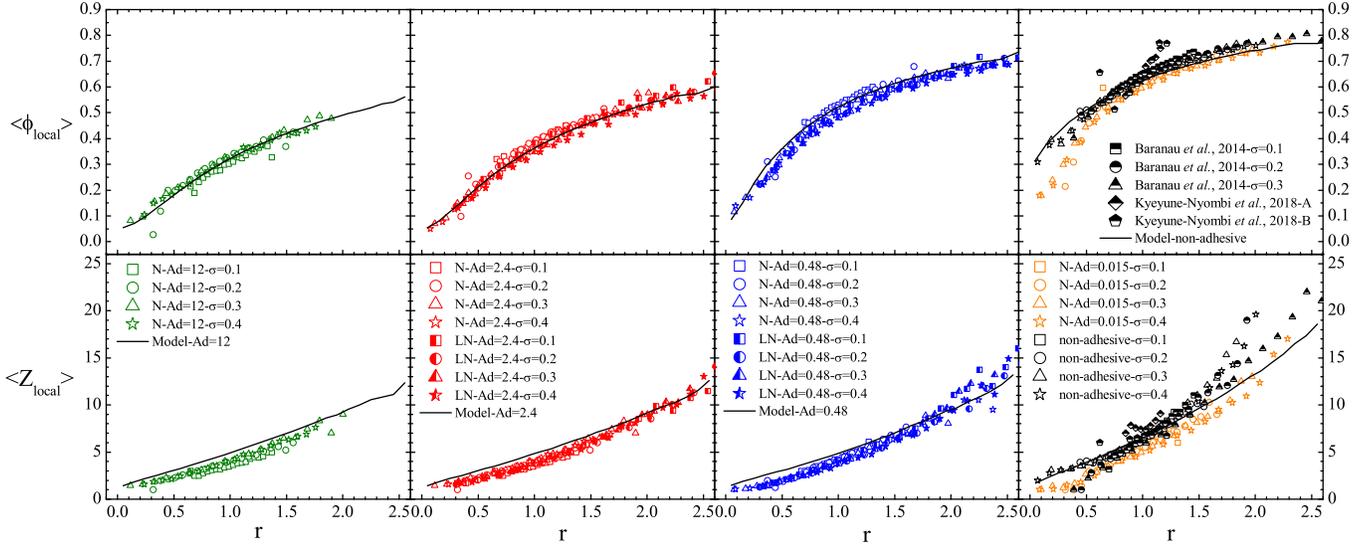}
\caption{\label{Fig2}(Colors online) Average local packing fraction (upper plot) and coordination number (lower plot) as a function of the normalized particle radius. The prefixes ``N'' and ``LN'' in the legend denote the Gaussian and lognormal size distributions, respectively. The lines are predictions from the modified model.}
\end{center}
\end{figure*}

Given the local packing fraction $\phi_{local}$ and local coordination number $Z_{local}$ of each particle in the packing, we are able to obtain the average value in terms of particle radius. We first divide the particle radius into many continuous intervals and then average the $\phi_{local}$ and $Z_{local}$ of the particles in each size interval. Figure \ref{Fig2} shows the average local packing fraction $<\phi_{local}>$ and coordination number $<Z_{local}>$ as a function of the normalized particle radius $r$. For comparison, packings of non-adhesive polydisperse spheres in both simulations and experiments from the literature are also incorporated \cite{Baranau14, Nyombi18}. We can see that the packing fraction moves to larger value as $Ad$ decreases, which agrees with the previous studies on packing of dry adhesive particles \cite{Yang00, Liu15, Liu17sm}. With a fixed $Ad$, $<\phi_{local}>$ grows consistently with the increase of $r$ and the $<\phi_{local}>-r$ profile falls on the same curve, which implies that there is no dependence of $<\phi_{local}>$ on the size variance $\sigma$ or both adhesive and non-adhesive particle packings. More interestingly, the $<\phi_{local}>-r$  profiles for Gaussian and lognormal size distributions seem to overlap, indicating that the location of the $<\phi_{local}>-r$ branch only depends on $Ad$. On the other hand, for different $Ad$, $<Z_{local}>$ seems to collapse onto a single curve for $Ad\geq 0.48$ regardless of the size distribution or size variance, while it develops to a different branch for the non-adhesive cases. Note that the $<\phi_{local}>-r$ and $<Z_{local}>-r$ profiles for very low $Ad$(=0.015) are quite close to those of non-adhesive cases, which approves a gradual transition from adhesive branch to non-adhesive one. By decreasing $Ad$, we are able to recover the non-adhesive cases, which confirms that $Ad$ is a well-defined parameter in describing the packings of adhesive particles \cite{Liu15, Liu17sm}. These observations indicate that there exist different size-topology relations in the very loose structures of adhesive particles.

\begin{figure}
\begin{center}
\includegraphics[width=7.5cm]{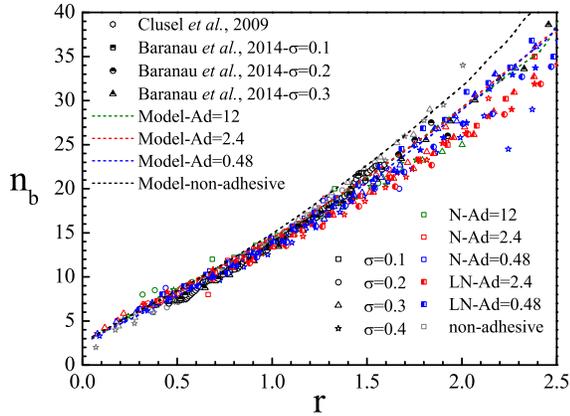}
\caption{\label{Fig3}(Colors online) average neighbour number $n_b$ as a function of the normalized particle size $r$. The solid lines are from the modified model.}
\end{center}
\end{figure}

From the geometrical view, the local packing structure can be constructed based on the relative location of the neighbouring particles. The more neighbours a single particle has, the more potential contacts there will be. Through the navigation map tessellation, we identify the neighbours around each particle in the packing. Figure \ref{Fig3} depicts the average neighbour number $n_b$ as a function of the normalized particle size $r$, including all our simulations as well as the results in literature. Intuitively, larger particles have more neighbours. However, it is fascinating that all the data points overlap onto the same curve and the $n_b-r$ relation seems to be quasilinear except for some deviations at large particles. This finding reveals that the determination of local neighbouring particles does not depend on the size span, or whether there is adhesion, which suggests that the local microstructure of each particle simply arises from the geometrical constraints.

Recently, a local model from the ``granocentric'' view for 3D polydisperse systems has been proposed to explain the size and neighbour topology in compressed emulsions and jammed non-adhesive granular matter \cite{Clusel09,Corwin10,Newhall11}. By randomly locating the neighbouring particles around a centric particle with some constraints, this model manages to construct the local geometry and well predicts the distributions of local packing fraction and coordination number, as well as the neighbour number. However, our simulation indicates that there exist different size-topology relations for dry adhesive particles that cannot be explained by the current model, even though they share the same $n_b-r$ profile. Therefore, a modified granocentric model is proposed to capture the geometry in packings of dry adhesive polydisperse spheres.

Starting from a centric particle with a radius selected from the prescribed size distribution, neighbouring particles are located around the centric particle one by one without overlapping, of which the size is also chosen from the same distribution as the centric particle. Each neighbouring particle is decided to be in contact with the centric particle with a probability $p_c$ and a surface-to-surface distance $d_s=0$, otherwise it is located with $d_s>0$ away from the centric particle. Then every placed neighbour $i$ occupies a solid angle $\Omega_i$ that depends on the radii of both the centric and neighbour particles, and $d_s$ \cite{Newhall11},
\be
\label{Eq4}
\Omega_i(r_c,r_p,d_s) = 2\pi (1-\frac{\sqrt{(r_c+r_{p,i}+d_s)^2-r_{p,i}^2}}{r_c+r_{p,i}+d_s}),
\ee
where $r_c$ denotes the radius of the centric particle. The total number of neighbours is determined when the sum of the solid angle occupied by each neighbour reaches a pre-set limit $\Omega^*_{tot}$. After deciding all the neighbours around the centric particle, the total cell volume $V_c$ can be calculated by summing up all the volume contributions from each neighbour, which is defined as a cone from the center of the centric particle to the hyperboloid surface determined by the navigation map. It should be noted that the total solid angle occupied by the neighbours are not equal to $4\pi$ due to the space between neighbours. However, the total solid angle occupied by the corresponding cones must be strictly equal to $4\pi$ when calculating the volume contribution, otherwise it will result in the underestimate of the total cell volume. In our modified model, the unoccupied solid angle, i.e. $4\pi - \Omega_{tot}$, is distributed to each neighbour proportionally to its solid angle, which is an important improvement from the previous model \cite{Newhall11}.

\begin{figure}
\begin{center}
\includegraphics[width=7.5cm]{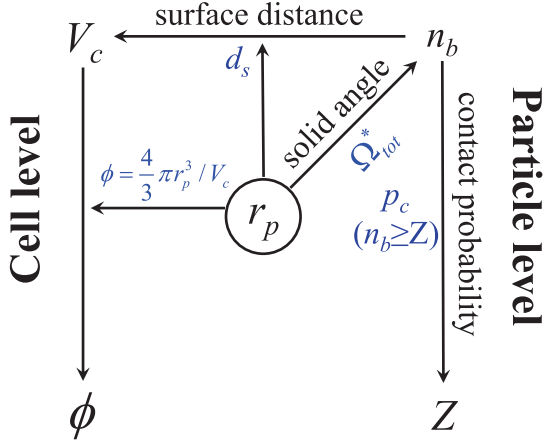}
\caption{\label{Fig4}(Colors online) The relationship between the local packing parameters in the model.}
\end{center}
\end{figure}

\begin{figure}
\begin{center}
\includegraphics[width=7.5cm]{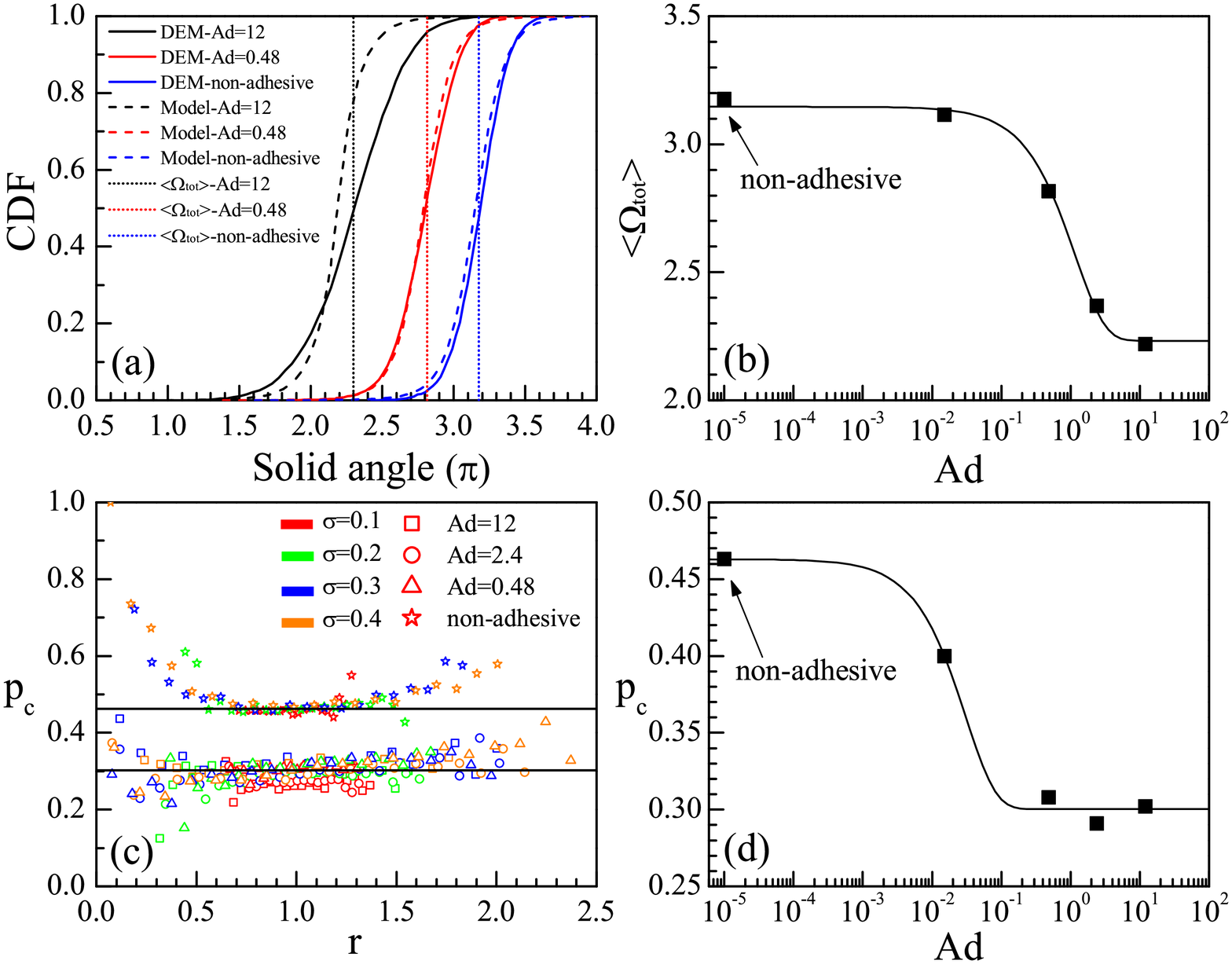}
\caption{\label{Fig5}(Colors online) (a) The cumulative distribution function of the total solid angle occupied by the neighbours for different $Ad$. The solid lines are the results obtained from the DEM simulation. The dashed lines are the predictions from the modified theory. The dotted lines represent the average total solid angle $<\Omega_{tot}>$. (b) The variation of average total solid angle $<\Omega_{tot}>$ as a function of the adhesion number $Ad$. The solid line is exponential fitting. (c) The contact probability $p_c$ for different $Ad$ and $\sigma$. The different shapes of the open symbols represent different $Ad$, while the color bars denote the size variance. (d) The variation of contact probability $p_c$ as a function of the adhesion number $Ad$. The solid line is exponential fitting. Note that in order to scatter the non-adhesive case (of which Ad is actually zero) on the semi-log plot, we manually set $Ad$ for the non-adhesive case with a very small value $10^{-5}$.}
\end{center}
\end{figure}

Figure \ref{Fig4} summarizes the procedure of how this model works and displays the relationship between the local parameters, which rely on three key parameters: (1) the total solid angle limit $\Omega^*_{tot}$, (2) the contact probability $p_c$, and (3) the surface-to-surface distance $d_s$. For the selection of $\Omega^*_{tot}$, the average total solid angle $<\Omega_{tot}>$ is suggested \cite{Newhall11}. It is indicated that $<\Omega_{tot}>$ does not depend on the centric particle radius and is approximately fixed at $<\Omega_{tot}> \approx 3.2\pi$ \cite{Clusel09, Newhall11} for jammed packings. However, our simulation shows that $<\Omega_{tot}>$ varies for adhesive particles (see Fig. \ref{Fig5}a), which becomes lower as $Ad$ decreases. Therefore, different values of $<\Omega_{tot}>$ must be applied for different $Ad$ in order to give a reasonable estimation of the number of neighbours. Figure \ref{Fig5}b shows the variation of $<\Omega_{tot}>$ as a function of $Ad$, which indicates that with the increase of $Ad$, $<\Omega_{tot}>$ gradually decreases from $3.2\pi$ to a very low value around $2.2\pi$. As for the contact probability $p_c$, it is also suggested that a fixed value of $p_c \approx 0.41$ is applied for non-adhesive particles \cite{Clusel09}. Nevertheless, $p_c$ decreases to around 0.3 for the adhesive particles in our simulation, as shown in Fig. \ref{Fig5}c, where there is still slight difference between different $Ad$. Similarly, the relationship between $p_c$ and $Ad$ is plotted in Fig. \ref{Fig5}d. It should be noted that the relationships between $<\Omega_{tot}>/p_c$ and $Ad$ can be described by simple exponential laws, as indicated by the solid lines in Fig. \ref{Fig5}b and \ref{Fig5}d. Therefore, the model parameters $<\Omega_{tot}>$ and $p_c$ can be readily accessed by estimating the $Ad$.

\begin{figure}
\begin{center}
\includegraphics[width=7.5cm]{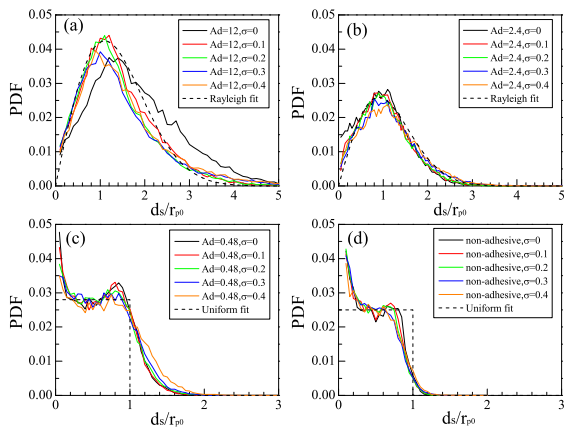}
\caption{\label{Fig6}(Colors online) The probability distribution function of surface-to-surface distance $d_s$ for (a) $Ad=12$, (b) $Ad=2.4$, (c) $Ad=0.48$, and (d) non-adhesive particles. The dashed lines in (a) and (b) are the Rayleigh fittings. The dashed lines in (c) and (d) are uniform distribution fittings.}
\end{center}
\end{figure}

Regarding the surface-to-surface distance $d_s$, Fig. \ref{Fig6} indicates that $d_s$ exhibits distinguished distributions for different $Ad$. For convenience, the remarkable peak at $d_s=0$, which corresponds to the contact between particles, has been removed. We can see from Fig. \ref{Fig6}a and \ref{Fig6}b that for $Ad=12, 2.4$, the probability distribution function (PDF) of $d_s$ grows a peak at $d_s=r_{p0}$, and decays gradually to zero after $d_s>4r_{p0}$. On the other hand, for $Ad=0.48$, it is almost constant in the range $d_s=0 \sim r_{p0}$ and the PDF drops drastically to zero after $d_s>r_{p0}$, which is very similar to that of non-adhesive particles (see Fig. \ref{Fig6}c and \ref{Fig6}d). Furthermore, it is worth noting that the PDFs of $d_s$ for the same $Ad$ are almost identical, indicating that they have no dependence on the size variance. Hence, in our modified model, we assume that the PDF of $d_s$ is described by the Rayleigh distribution for $Ad>0.48$ (dashed lines in Fig. \ref{Fig6}a and \ref{Fig6}b) based on a best fitting, while it is approximated by a uniform distribution between $d_s=0$ and $d_s=r_{p0}$ for $Ad\leq 0.48$.

Given the particle size distribution $P(r_p)$, it is feasible to write the master equations to obtain the PDF of the number of neighbour, and further the PDFs of contact number and local volume fraction \cite{Clusel09,Corwin10,Newhall11}. The master equations for our modified model keep the same forms as those in the previous model \cite{Newhall11}. The improvements lie in the extension of the local parameters, which relates the model parameters to the interparticle adhesion. However, to solve these equations analytically are very impractical. Therefore, we resort to employ efficient Monte Carlo simulations to create numerous local cells based on the above procedure and thus accurately obtain the full distributions of the neighbour number, the contact number, as well as the local packing fraction. Before the MC simulation, appropriate values of the three model parameters must be determined according to $Ad$. For instance, $\Omega^*_{tot}=2.2\pi$, $p_c=0.3$ and a Rayleigh distribution of $d_s$ with the scale parameter equal to 1.28 are used for $Ad=12$. Then the local cells can be generated with the following algorithm.

({\it i}) Select a centric particle's radius, $r_c$, from the prescribed distribution, $P(r_p)$. Then select $m$ potential neighbouring particles' radii, $r_{p,i}(i=1,2,3,...,m)$, from the same distribution $P(r_p)$, where $m$ must be greater than the maximum number of expected neighbouring particles (for example $m=50$).

({\it ii}) Decide whether these $m$ potential neighbouring particles are in contact with the centric particle with the contact probability $p_c$. Generate a random number between 0 and 1 for each particle. If this random number is lower than or equal to $p_c$, then this particle is in contact with the centric particle. Otherwise, it is not. After that, set the first $Z_{min}=2$ potential neighbours to be contacting neighbours. This is because the minimum mean coordination number for adhesive particles are 2 according to the adhesive loose packing (ALP) conjecture \cite{Liu15, Liu17sm}.

({\it iii}) Assign the surface-to-surface distance $d_s$ to each potential neighbouring particle. For those neighbours that are not in contact with the centric particle, select $d_s$ from the prescribed distribution (either Rayleigh or uniform distribution in this study). For those contact neighbours, $d_s=0$.

({\it iv}) Calculate the solid angle each potential neighbour occupy based on Eq. \ref{Eq4}.

({\it v}) Determine the actual number of neighbours by calculating the cumulative sum of the solid angle. If the cumulative sum of first $n_b$ potential neighbours is less than the limit $\Omega^*_{tot}$, and that of $n_b+1$ potential neighbours is greater than the limit $\Omega^*_{tot}$, then the actual neighbour number is $n_b$. Note that half the time, the next potential neighbour is also included, which guarantees that the average total solid angle is close to the limit $\Omega^*_{tot}$. After deciding the actual neighbour number, the actual coordination number equals the number of contacting particles among the actual neighbours.

({\it vi}) Calculate the total solid angle $\Omega_{tot}$ of the actual neighbours, $\Omega_{tot}=\sum\limits_{i=1}^{n_b}\Omega_i$. Distribute the remaining solid angle to each neighbour proportionally to its solid angle, $\Delta \Omega_i=(4\pi-\Omega_{tot})\frac{\Omega_i}{\Omega_{tot}}$. Then the real solid angle of each neighbour for calculating the volume contribution is $\Omega_i+\Delta \Omega_i$.

({\it vii}) Calculate the cell volume by summing up all the contribution from each neighbour. First, calculate the angle of the cone defined by the real solid angle \cite{Newhall11}, 
\be
\label{Eq5}
\theta_c = \arccos (1-\frac{\Omega_i+\Delta \Omega_i}{2\pi}).
\ee
Second, determine the location of the hyperbolic sheet cap defined by the navigation map in polar coordinates, $(\hat{r},\hat{\theta})$. Taking $\hat{\theta}=0$ as the line connecting the centers of the centric particle and the neighbour, the hyperbolic sheet's location is obtained as \cite{Newhall11}
\be
\label{Eq6}
\hat{r}(\hat{\theta}) = \frac{r_{p,i}^2+(r_c-r_{p,i})^2}{2r_{p,i}\cos\hat{\theta}-2(r_c-r_{p,i})},
\ee
which is rotationally symmetric about $\hat{\theta}=0$. Finally, integrate the volume in spherical coordinates from $\hat{r}=0$ to the cap $\hat{r}(\hat{\theta})$ defined above, over $\hat{\theta}$ from $\hat{\theta}=0$ to $\hat{\theta}=\theta_c$, and over $\varphi$ from 0 to $2\pi$ \cite{Newhall11},
\be
\begin{split}
\label{Eq7}
V_{c,i} &= \frac{\pi(d_{s,i}+2r_c)^3(d_{s,i}+2r_{p,i})}{24(d_{s,i}+r_c+r_{p,i})} \\
&\times \left\{\frac{(d_{s,i}+2r_{p,i})^2}{[r_{p,i}-r_c+(d_{s,i}+r_c+r_{p,i})\cos\theta_c]^2}-1 \right\}.
\end{split}
\ee
Then the cell volume equals $V_c=\sum\limits_{i=1}^{n_b}V_{c,i}$.

\begin{figure}
\begin{center}
\includegraphics[width=7.5cm]{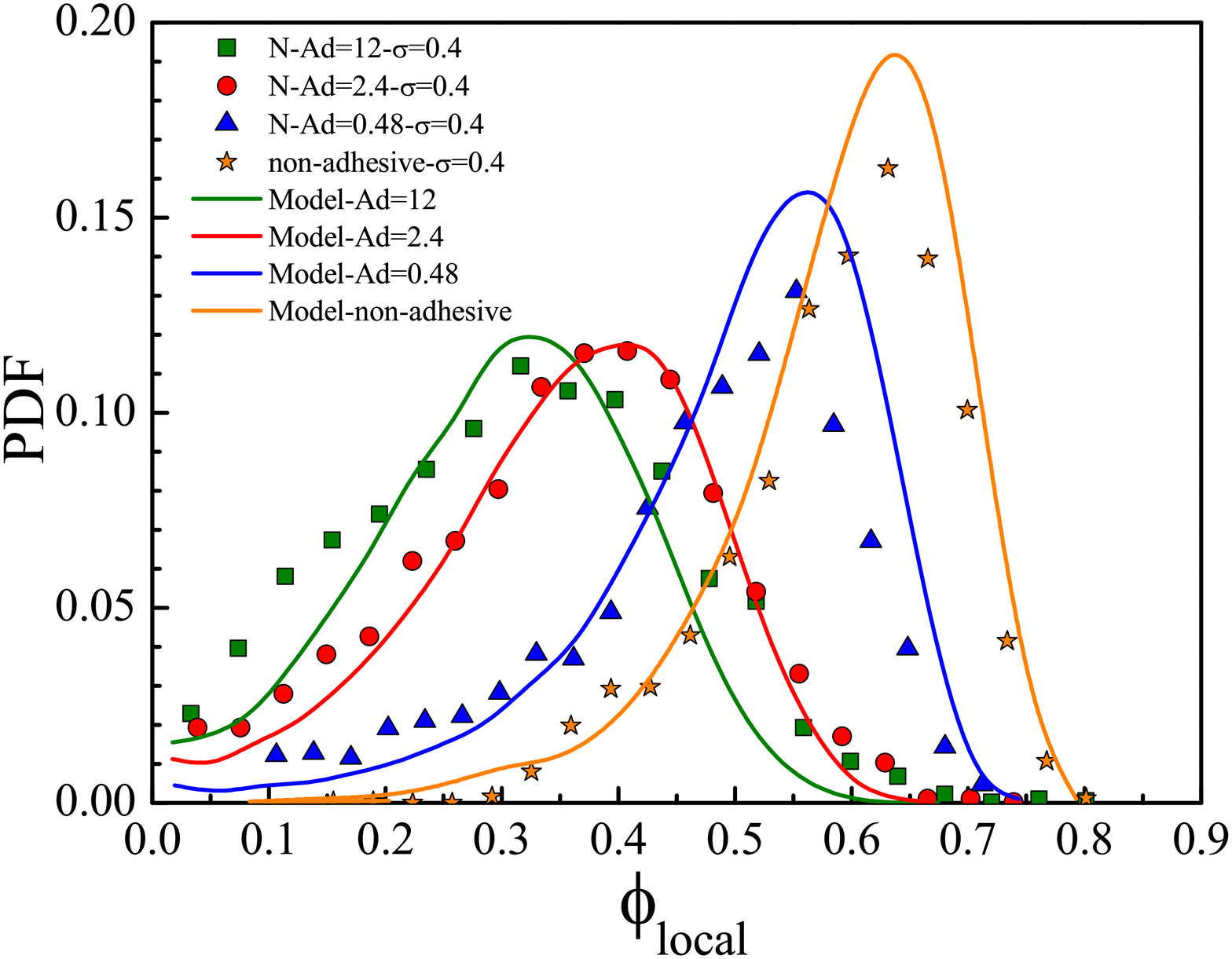}
\caption{\label{Fig7}(Colors online) Distribution of the local packing fraction with $\sigma=0.4$ for different $Ad$. The points are the DEM simulation results and the solid lines are the theoretical predictions obtained from the MC simulation based on the modified model.}
\end{center}
\end{figure}

After the seven steps above, a local cell is constructed with the neighbour number, contact number and cell volume all determined. It should be noted that the algorithm can be efficiently implemented with the matrix form, with which millions of cells can be generated simultaneously. Then the PDF of all the local packing properties can be subsequently obtained. Figure \ref{Fig7} shows the PDF of the local packing fraction from both DEM simulation and the model, where fairly good agreement is reached. Obviously, as $Ad$ decreases, the global packing fraction increases, which resembles the non-adhesive cases when $Ad\ll 0.48$. Furthermore, theoretical predictions of the $<\phi_{local}>-r$, $<Z_{local}>-r$, $n_b-r$ and solid angle profiles are also presented in both Fig. \ref{Fig2}, Fig. \ref{Fig3} and Fig. \ref{Fig5}a, respectively, which all agree well with our DEM simulation and the literature. Not only do we well describe the size-topology relations for dry adhesive particles, but also recover the results of non-adhesive jammed packings, which indeed approves the validity of our modified model.

\section{Conclusion}

As a summary, we systematically investigate the random packing of dry adhesive microspheres covering two different size distributions and a wide range of size span. The influence of interparticle adhesion is well characterized by a dimensionless adhesion number, $Ad$. The geometry of the packing structure is tilled through the navigation map tessellation. A unifying quasilinear relation between the average neighbour number and the normalized particle size is revealed, regardless of the size distribution or the interparticle adhesion. This finding confirms that the number of neighbouring particles in a local cell configuration are simply determined by the geometry that depends on the relative size ratio of the centric particle. Generally, the larger the centric particle is, the more neighbours it has. Moreover, the regressions of both the average local packing fraction and coordination number as a function of the normalized particle size are found to be independent of the size variance. With the decrease of the adhesion number (decrease in the adhesion strength), both the $<\phi_{local}>-r$ and $<Z_{local}>-r$ profiles undergo gradual transitions from adhesive branches to non-adhesive branch. More importantly, we didn't find any obvious distinctions between the results of the two different size distributions, i.e. Gaussian and lognormal distributions. This implies that size-topology relation might be a unique feature in polydisperse particle system, which needs to be further validated with other size distributions. Such adhesion induced size-topology relations found in our work are well explained by a modified granocentric model, where the model parameters are further extended beyond the non-adhesive case and their novel relationships with the adhesion number are established. Our findings provide fundamental knowledge on the geometry of loose polydisperse particle system, which may shed light on the understanding of the underlying mechanisms relating to the formation of adhesive assembly.

\section{Acknowledgements}
This work has been funded by the National Natural Science Foundation of China (No. 51725601 and No. 51390491). S. Q. Li is grateful to Prof. Jeff Marshall at University of Vermont, Prof. Hern\'a n Makse at City College of New York, Dr. Guanqing Liu and Dr. Mengmeng Yang at Tsinghua University for helpful discussions. W. Liu acknowledges the support from the Engineering and Physical Sciences Research Council (EPSRC, Grants No: EP/N033876/1).

\bibliography{refpre}
\bibliographystyle{apsrev4-1}

\end{document}